\documentclass[]{spiejour}  

\usepackage{epsfig}
\usepackage{amsmath,amssymb,mathrsfs}

\title{Field enhancement at metallic interfaces due to quantum confinement}

\author{Z. Fatih {\"O}zt{\"u}rk,\supit{a,b} Sanshui Xiao,\supit{a} Min Yan,\supit{a,c} Martijn Wubs,\supit{a} Antti-Pekka Jauho,\supit{d,e} and N. Asger Mortensen\supit{a}
\skiplinehalf
\supit{a} Department of Photonics Engineering, Technical University of Denmark, DK-2800 Kongens Lyngby, Denmark \\
\supit{b} Istanbul Technical University, Istanbul, Turkey \\
\supit{c} Laboratory of Photonics and Microwave Engineering, Royal Institute of Technology, Sweden \\
\supit{d} Department of Applied Physics, Aalto University, Finland \\
\supit{e} Department of Micro and Nanotechnology, Technical University of Denmark, DK-2800 Kongens Lyngby, Denmark
}

\authorinfo{Send correspondence to N.A.M.: asger@mailaps.org}

\begin{document}
  \maketitle

\begin{abstract}
We point out an apparently overlooked consequence of the boundary
conditions obeyed by the electric displacement vector at air-metal
interfaces: the continuity of the normal component combined with the
quantum mechanical penetration of the electron gas in the air
implies the existence of a surface on which the dielectric function
vanishes. This, in turn, leads to an enhancement of the normal
component of the total electric field. We study this effect for a planar metal surface, with the inhomogenous electron density accounted for by a Jellium model. We also illustrate the effect
for equilateral triangular nanoislands via numerical solutions of
the appropriate Maxwell equations, and show that the field
enhancement is several orders of magnitude larger than what the
conventional theory predicts.
\end{abstract}


\maketitle

\section{Introduction}

Electromagnetic field enhancement in nanoplasmonic structures plays
an important role in a number of applications such as nano
antennas~\cite{Muhlschlegel:2005,Novotny:2007}, single-photon
emitters~\cite{Chang:2007}, and surface-enhanced spectroscopy
techniques in general~\cite{Moskovits:1985}, including
surface-enhanced Raman spectroscopy (SERS)~\cite{Fleischmann:1974}.
Field localization is the key to understanding and further enhancing
the resolution of SERS, and the use of highly engineered surface
plasmonic structures has been envisioned~\cite{Lal:2007,Xiao:2008}. Strong field enhancement has so far been inherently connected to singularities~\cite{Luo:2010}.

It is commonplace to study the plasmonic response using Maxwell's
equations with the bulk dielectric function of the metal, with an
abrupt change at the air-metal interface~\cite{Maier:2007}. Though clearly neglecting
surface and confinement effects for the electrons in the metal, this
is a celebrated approximation which predicts the existence of
surface plasmons forming the heart of nanoplasmonics. However, when
samples involve true nano-scale feature-sizes, the validity of this
approximation is challenged and new phenomena, such as non-local response and spatial
dispersion~\cite{Lindhard:1954,Boardman:1976,Dasgupta:1981,Chang:2006,abajo:2008,McMahon:2009,McMahon:2010a,McMahon:2010b} and inhomogeneous electron densities due to quantum confinement~\cite{Zuloaga:2009,PerezGonzalez:2010} are expected to play an important
role.

In nanoscale metallic structures the electrons are subject to
quantum confinement, which leads to a number  of spectacular
effects, such as the quantized conductance of atomic quantum-point
contacts~\cite{Yanson:1998}, Friedel oscillations and confinement of
electrons to quantum corrals on noble metal
surfaces~\cite{Crommie:1993}, and a size-dependent structure of
nanocrystals~\cite{Lauritsen:2007}. Light emission from STM has been used to reveal plasmonic properties of electrons confined
to triangular islands on surfaces of noble
metals~\cite{schull:2008}. All these studies clearly show that, at
least at low temperatures, quantum confinement plays an important
role leading to inhomogeneous electron densities at interfaces that differ from the piecewise constant bulk values.

\section{From bulk approximation to a local density approximation}

The qualitative optical consequences of an inhomogeneous electron density
can be explored within Lindhard's \emph{``local density
approximation''} for the dielectric
function~\cite{Lindhard:1953,Nitta:1993}
\begin{equation}\label{eq:Lindhard}
\epsilon({\boldsymbol r},\omega)=
1-\frac{\omega_p^2}{\omega(\omega+i\gamma)} \Pi({\boldsymbol
r})
\end{equation}
where $\Pi({\boldsymbol r})=n({\boldsymbol r})/n_0$ accounts for the
density variations relative to the bulk density of electrons
$n_0=m\omega^2_p/e^2$, with $\omega_p$ being the bulk plasma
frequency. To mimic the Ohmic damping we have included a damping
rate $\gamma$ and in the optical regime $\gamma \ll \omega$~\cite{Mermin:1970}. While several studies
have considered the consequences of local density variations for the
surface-plasmon dispersion relation, the overall effects have been
reported to be minor and for planar surfaces the surface-plasmon
resonance at $\omega=\omega_p/\sqrt{2}$ appears unaffected by the
detailed density profile at the metal-air
surface~\cite{Boardman:1975}. However, it remains an open question
to which extent the detailed field distribution will be affected by
a smoothly varying electron density in contrast to the discontinuous
drop in the electron density assumed by the bulk approximation.

Zero-index phenomena in metamaterials are receiving increasing attention~\cite{garcia:2002,Silveirinha:2006,cheng:2007,Liu:2008,Edwards:2008,Pollard:2009}, and the recent results by Litchinitser \emph{et al.}~\cite{Litchinitser:2008} show
that for a metamaterial with the effective constitutive coefficients
varying linearly, $\epsilon(z)\propto z$ and $\mu(z)\propto z$, a
pronounced field enhancement will occur at $z=0$ where
$\epsilon=\mu=0$. In this paper we show that plasmonic structures
employing Nature's own metals may support large field enhancement at
the zero-$\epsilon$ surface appearing in the vicinity of the
air-metal interface, where the real part of the polarization of the dilute electron
gas will exactly balance the polarization of the vacuum background. As shown below, the
particular position of the zero-$\epsilon$ surface depends on the
frequency $\omega$ of the incident radiation, and may be either in
air or in the metal, depending on the local geometry.

\section{Phenomenology of field enhancement}
The basic mechanism behind the field
enhancement can be understood from the continuity condition for the
electrical displacement field, which stipulates that
${\boldsymbol n}\cdot {\boldsymbol D} = \epsilon ({\boldsymbol
n}\cdot {\boldsymbol E})$ be continuous (here $\boldsymbol n$ is the normal vector of a
surface element). At the zero-$\epsilon$ surface the normal
component $E_\perp = {\boldsymbol n}\cdot {\boldsymbol E}$ thus has
to diverge to keep the product $\epsilon E_\perp $  finite.

From Eq.~(\ref{eq:Lindhard}) it follows that
in the regions where  $\mathrm{Re}(\epsilon)$ vanishes, there will
still be a small, but finite imaginary part:
$\mathrm{Im}(\epsilon)_{\mathrm{Re}(\epsilon)=0}= \gamma/\omega$.
For an  electrical field ${\boldsymbol E}_0$ incident at an angle
$\theta$ we thus expect that the normal component ${\boldsymbol
n}\cdot {\boldsymbol E}_0 = E_0 \sin\theta$ is enhanced by a factor
proportional to $1/\gamma$ on the zero-$\epsilon$ surface. The
intensity enhancement can then be estimated as
\begin{equation}\label{scaling}
\big|{\boldsymbol E}\big|^2\big/\big|{\boldsymbol E}_0\big|^2 \propto
\sin^2\theta/\gamma^2
\end{equation}
We emphasize that this mechanism of field-enhancement is generic and
robust as it only depends on the assumed continuous negative-to-positive variation of the dielectric function at the metal-air interface, in combination with a Maxwell boundary
condition~\cite{Sondergaard:2007}, rather than on the details of the
excitation of strongly localized surface-plasmon resonances with
pronounced sample-to-sample fluctuations.

\section{Example 1: Jellium model for the metal surface}

\begin{figure}[t!]
\begin{center}
\epsfig{file=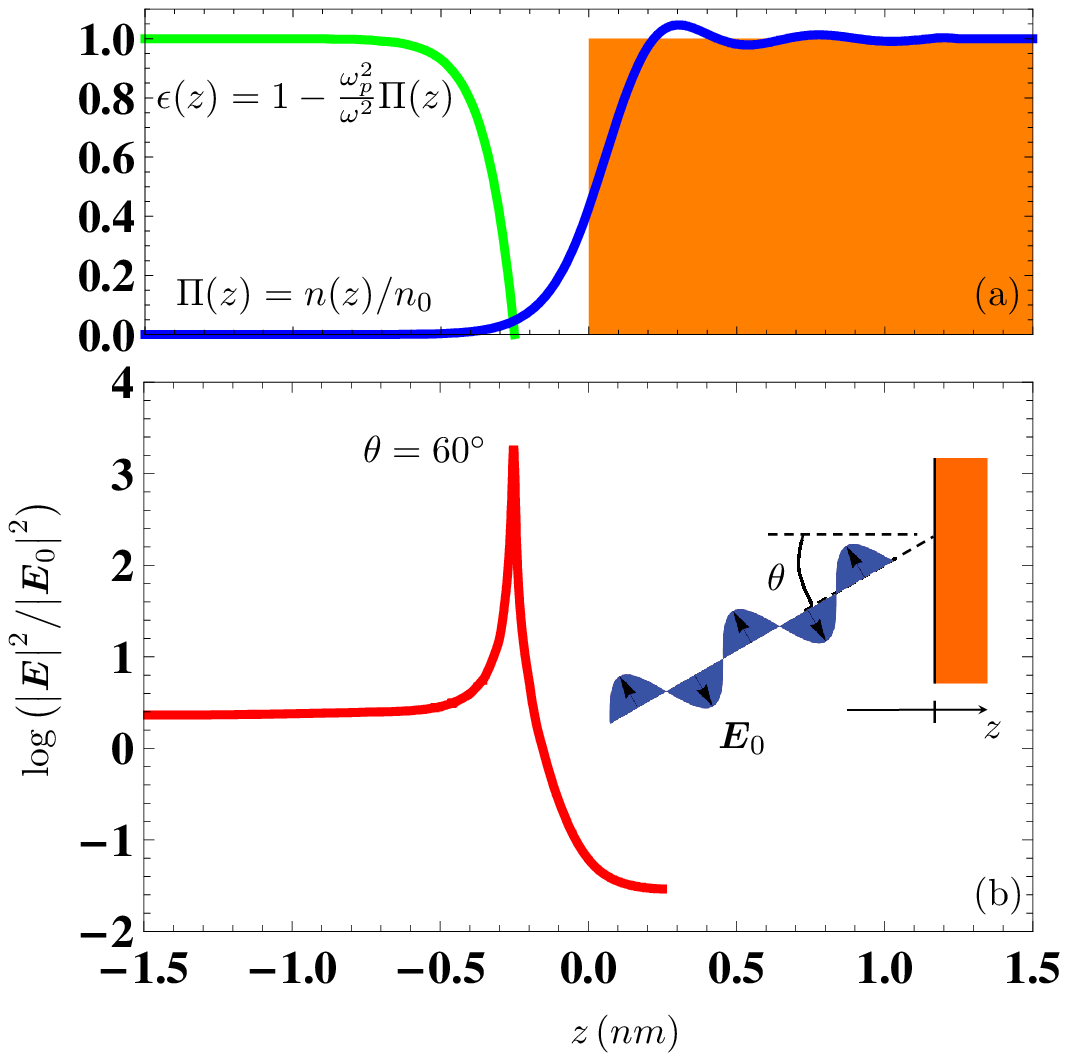, width=0.8\linewidth,clip}
\caption{(a) The blue line shows the position dependent electron density within the Jellium model ($r_s/a_0=2$). The green line shows the corresponding derived dielectric function. (b) The field enhancement for a plane wave incident
on the surface at an angle of $\theta=60^\circ$, for
$\lambda=633~{\rm nm}$, $\omega_p=1.37\times 10^{16}\,{\rm
rad}/{\rm s}$, and $\gamma = 1.0\times 10^{14}\,{\rm rad}/{\rm s}$.
 }
\label{fig1}
\end{center}
\end{figure}

The low-temperature density profile for Jellium is a function of
$r_s/a_0$ only, with the radius of the free-electron sphere $r_s=(3/4\pi n_0)^{1/3}$ and $a_0=\hbar^2/me^2$ being the Bohr radius.
The position-dependent electron density may be obtained self-consistently
with the aid of density-functional theory~\cite{Lang:1970},
thus taking into account both exchange and correlation effects in the inhomogeneous electron gas. In Ref.~\cite{Lang:1970}, results are tabulated for a range of values of $r_s/a_0$.
The resulting density $n(z)$ is shown in
Fig.~\ref{fig1}(a)  for $r_s/a_0=2$, a typical value for gold or
silver. As seen, $\Pi(z)=n(z)/n_0$ varies continuously on the length
scale of the Fermi wavelength rather than exhibiting a discontinuous
drop at the surface of the metal ($z=0$). In the metal ($z>0$),
there are Friedel oscillations near the surface, while for $z<0$ the
decay of the density is caused by the finite value of the work
function $W$, which allows electron wave functions to penetrate into
the classically forbidden region.
According to Eq.~(\ref{eq:Lindhard}), the dielectric function will
inherit the Friedel oscillations in the metal, as well as approach
the value in vacuum  outside the metal surface. As a consequence,
the dielectric function passes through zero just outside the metal
surface [green line in Fig.~\ref{fig1}(a)].
Similar spatial variations of the dielectric function have been
reported in more elaborate studies of e.g. semiconductor quantum
dots~\cite{Cartoixa:2005} and recently also in doped semiconductors~\cite{Feigenbaum:2010}.

\begin{figure}[t!]
\begin{center}
\epsfig{file=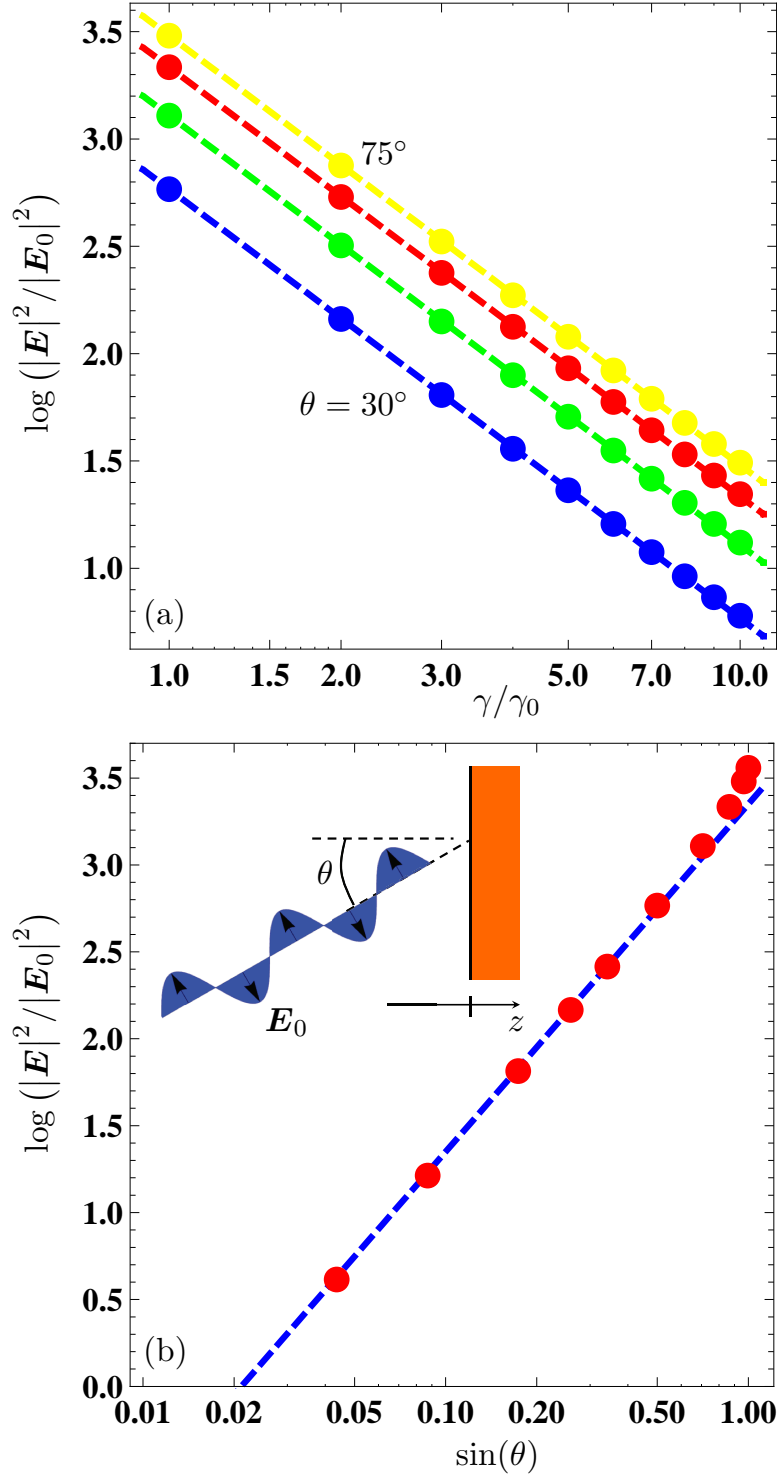, width=0.8\linewidth,clip}
\caption{(a) Maximal field enhancement as a function of damping for parameters of Fig.~\ref{fig1}(b). The dashed lines indicate a $\gamma^{-2}$ dependence. (b) The peak height as a function of incident angle for $\gamma = 1.0\times 10^{14}\,{\rm rad}/{\rm s}$. The dashed line indicates a $\sin^2\theta$ dependence,
Eq.(\ref{scaling}).
 }
\label{fig2}
\end{center}
\end{figure}

In the case of $\lambda=2\pi c/\omega= 633~{\rm nm}$ the
zero-$\epsilon$ surface appears at a distance of 0.25~nm from the
surface  ($z=0$). Fig.~\ref{fig1}(b) illustrates the field
enhancement in the case of a plane wave incident at an angle of
$\theta=60^\circ$. The numerical results were obtained by solving
the wave equation
\begin{equation}
\nabla\times\nabla\times {\boldsymbol E}({\boldsymbol
r})=\epsilon({\boldsymbol r},\omega)\frac{\omega^2}{c^2} {\boldsymbol
E}({\boldsymbol r})
\end{equation}
with the aid of a finite-element method employing an adaptive mesh
algorithm (Comsol MultiPhysics). A comparison with $\epsilon(z)$ in
Fig.~\ref{fig1} (a)  shows that the field enhancement indeed occurs
at the point where $\epsilon=0$. We emphasize that this phenomenon
has no counterpart within the classical bulk approximation of the
dielectric function. For a realistic value of the damping
($\gamma=1\times 10^{14}\,{\rm rad}/{\rm s}$)  the intensity may be
enhanced by more than 3 orders of magnitude, which suggests that in
a SERS experiment the Raman signal could be enhanced by 6 orders of
magnitude~\cite{Garciavidal:1996} without the need for any geometry-induced localized
resonances. To further test the predictions of Eq.~(\ref{scaling})
we have explored the dependence on the angle of incidence as well as
the particular value of the plasmon damping. Fig.~\ref{fig2} (a)
illustrates how the peak height depends on damping.
The enhancement is limited by the plasmonic damping with the slope
of the dashed lines being in full accordance with the power (-2) of
$\gamma$ in Eq.~(\ref{scaling}). Panel (b) illustrates the angle
dependence with the dashed line showing the $\sin^2\theta$
dependence of Eq.~(\ref{scaling}), indicating a pronounced
enhancement for oblique incidence.

\section{Example 2: A metallic nanostructure}

\begin{figure}[t!]
\begin{center}
\epsfig{file=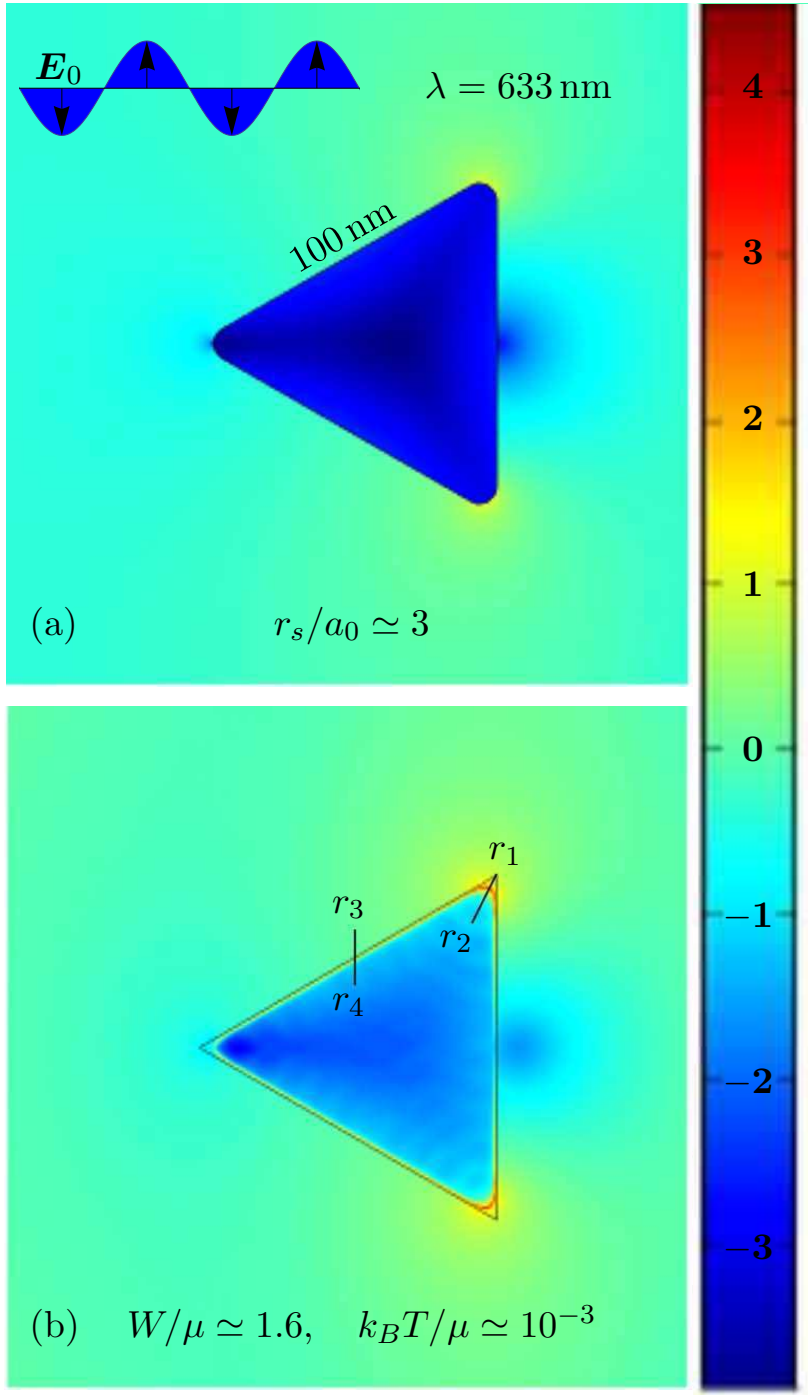, width=0.8\linewidth,clip}
\caption{(a) Intensity enhancement (log scale) for a wave incident from the left on a metallic equilateral triangle, computed with the bulk dielectric function, but with rounded corners (radius of 5~nm). (b) Field enhancement for the inhomogeneous density. Field profiles along the lines $r_1\rightarrow r_2$ and $r_3\rightarrow r_4$ are shown in Fig.~\ref{fig4}. } \label{fig3}
\end{center}
\end{figure}

\begin{figure}[t!]
\begin{center}
\epsfig{file=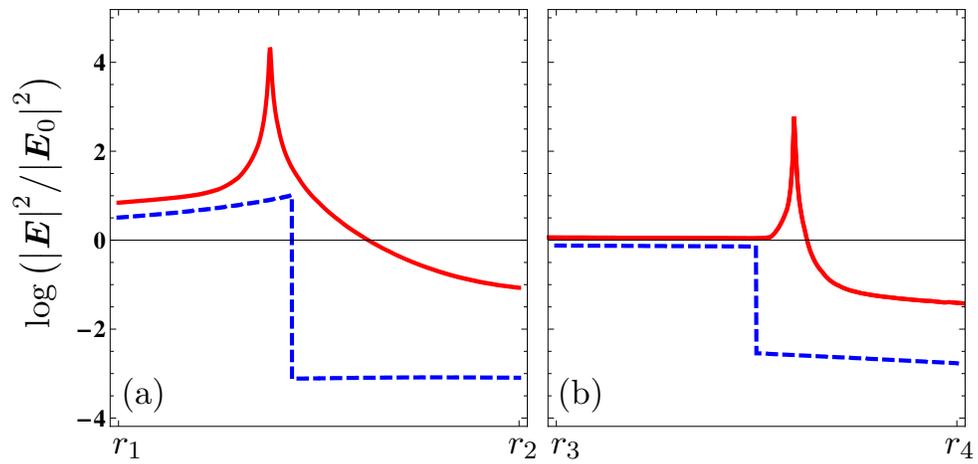, width=\linewidth,clip}
\caption{Field enhancement along the lines defined in Fig. 3. (a) $r_1\rightarrow r_2$. (b) $r_3\rightarrow r_4$. Note that the integrated density is the same for the two structures in Fig.~\ref{fig3}, thus causing slightly different densities and field suppression inside the metal structures (compare dashed and solid lines for the inhomogeneous and homogeneous cases, respectively).} \label{fig4}
\end{center}
\end{figure}

We next explore the
field enhancement in the presence of strong confinement of electrons
in metallic nanostructures. Here we limit ourselves to a model of
non-interacting electrons in wire geometries with nanoscale cross
sections $\Omega$ so that the problem effectively is
two-dimensional. The energy states of the electron system are then
given by ${\mathscr E}(\kappa)=\tfrac{\hbar^2}{2m}\kappa^2+{\mathscr
E}_\nu$ with the transverse energy components being quantized and
governed by a two-dimensional Schr{\"o}dinger equation
\begin{equation}\label{eq:schrodinger}
\left[-\frac{\hbar^2}{2m}\nabla_{xy}^2+V(x,y)\right]\psi_\nu(x,y)={\mathscr E}_\nu\psi_\nu(x,y).
\end{equation}
For the confinement potential $V$ we adjust the height in accordance
with the work function of the metal $W=V_0-\mu$, where $\mu$ is the chemical potential. In equilibrium the
states are populated according to the Fermi--Dirac distribution
function and  we find
\begin{equation}\label{eq:Jonquiere}
\Pi(x,y) \equiv\frac{n(x,y)}{n_0} =  \frac{\sum_\nu
P\left(\frac{{\mathscr
E}_\nu-\mu}{k_BT}\right)A\big|\psi_\nu(x,y)\big|^2}{\sum_\nu
P\left(\frac{{\mathscr E}_\nu-\mu}{k_BT}\right)}
\end{equation}
where $A=\int_\Omega dx dy$ is area and $P(x)= {\rm
Li}_{1/2}[-\exp(-x)]$ with ${\rm Li}_n(x)$ being Jonqui{\'e}re's
polylogarithm function. Equilateral triangular nano islands have
been the subject of numerous
investigations~\cite{Sundaramurthy:2006,Nelayah:2007} and in
Fig.~\ref{fig3} we consider field enhancement in such a structure.
Fig.~\ref{fig3}(a) shows results within the bulk approximation,
where an artificial rounding of the corners has been added to
circumvent the effect of an otherwise divergent electrical
field~\cite{Xiao:2008}. Clearly, field enhancement is modest since
no pronounced surface plasmon resonances are excited.
Fig.~\ref{fig3}(b) on the other hand is based on the spatially continuously varying dielectric function of Eqs.~(\ref{eq:Lindhard}) and (\ref{eq:Jonquiere}) and shows a pronounced field
enhancement along the zero-$\epsilon$ surface. We also note that
while the geometry has arbitrarily sharp corners, the electron
density and dielectric function themselves are still smooth, showing
that quantum confinement provides a built-in smoothing even in the
presence of sharp geometrical features. Fig.~\ref{fig4}(a) and (b)
show in more detail the field enhancement along the lines indicated in Fig.~\ref{fig3}(b). For the results in Fig.~\ref{fig4}(b), the peak height clearly matches the one-dimensional results in
panel (b) of Fig.~\ref{fig2} for an incident angle of
$\theta=30^\circ$. Similarly, the result within the bulk
approximation resembles typical results, see
e.g.~Ref.~\cite{Sondergaard:2007}, i.e. the field enhancement is more modest. In Fig.~\ref{fig4}(a), the incident
wave propagates almost parallel to the zero-$\epsilon$ surface ($\theta\sim
90^\circ$), thus causing a peak that is even higher at the two
right-most corners of the triangle. On the other hand, at the left-most
corner the wave has normal impact on the zero-$\epsilon$ surface
($\theta=0^\circ$) so that no pronounced field enhancement is
observed.  It is interesting to note that the enhancement may occur
either outside [as in Fig.~\ref{fig1}(b)] or inside [as in Fig.~\ref{fig3}(b)]
the metal, depending on the local geometry and the frequency relative to the plasma frequency.

\section{Conclusions}

We predict that a large non-resonant field
enhancement may occur at zero-epsilon surfaces near the air-metal
interface. Theoretically, the enhancement occurs in a wide class of models where the discontinuous jump of the dielectric function of the usual bulk description of at an air-metal interface is replaced by a continuous function $\varepsilon({\bf r},\omega)$ that has the bulk values as limiting values away from the interface. We discussed two microscopic models for such a continuous variation at the interface.  The predicted field enhancement has therefore no counterpart in the classical bulk treatment
of plasmonic field enhancement, though it is intimately related to
the familiar boundary condition for the electrical displacement
field. In particular, for light impinging on an interface we discussed the dependence of the field enhancement on the angle of incoming light and on the damping in the metal. Furthermore, the enhancement constitutes another mechanism for the enhancement of Raman signals at metal interfaces (SERS).

Interestingly, Feigenbaum \emph{et al.} have recently reported their experimental observation of zero-index driven field enhancement at the surface of conducting oxides (i.e. with lower electron density compared to good metals)~\cite{Feigenbaum:2010}. We emphasize that the decay of the electron density into vacuum (or more generally into a dielectric) could have important implications also for nano-gap
structures (such as the gap formed between the corners of two nearby
triangles~\cite{Sundaramurthy:2006}), where quantum tunneling of
electrons will introduce a small but finite electron density inside
the gap, thus also modifying the local dielectric function and the ability to support short-circuiting currents~\cite{PerezGonzalez:2010}.

\section*{Acknowledgments}

We thank Peter Nordlander, Natalia M. Litchinitser, and Vladimir M. Shalaev for useful discussions. This work is financially supported by the Danish Council for
Strategic Research through the Strategic Program for Young
Researcher (grant no: 2117-05-0037), the Danish Research Council for
Technology and Production Sciences (grants no: 274-07-0080 \&
274-07-0379), as well as the FiDiPro program of the Finnish Academy.

\newpage

\bibliographystyle{spiejour}

\newpage

\end{document}